\documentclass{bioinfo}
\copyrightyear{2021} \pubyear{2021}

\usepackage{subfiles}
\access{Advance Access Publication Date: Day Month Year}
\appnotes{Manuscript Category}
\usepackage{placeins}

\usepackage{color, colortbl}
\usepackage{xcolor}
\definecolor{blue}{rgb}{1.,0,0}
\usepackage[final]{changes}
\definecolor{blue}{rgb}{0,0,1.}
\definecolor{Gray}{gray}{0.9}

\definecolor{myblue}{rgb}{0, 0, 0.7}

\begin{document}
\firstpage{1}

\subtitle{Subject Section}

\title[TITAN: T Cell Receptor Specificity Prediction]{TITAN: T Cell Receptor Specificity Prediction with Bimodal Attention Networks}
\author[Weber \textit{et~al}.]{Anna Weber\,$^{\text{\sfb 1,2,}*}$, Jannis Born\,$^{\text{\sfb 1,2}}$ and Mar\'ia Rodr\'iguez Mart\'inez\,$^{\text{\sfb 1,}}$}
\address{$^{\text{\sf 1}}$IBM Research, Zurich, Switzerland and \\
$^{\text{\sf 2}}$ETH Zurich, Switzerland.}

\corresp{$^\ast$To whom correspondence should be addressed.}

\history{Received on XXXXX; revised on XXXXX; accepted on XXXXX}

\editor{Associate Editor: XXXXXXX}

\abstract{
\textbf{Motivation:} The activity of the adaptive immune system is governed by T-cells and their specific T-cell receptors (TCR), which selectively recognize foreign antigens. Recent advances in experimental techniques have enabled sequencing of TCRs and their antigenic targets (epitopes), allowing to research the missing link between TCR sequence and epitope binding specificity. Scarcity of data and a large sequence space make this task challenging, and to date only models limited to a small set of epitopes have achieved good performance. Here, we establish a \added{$k$-nearest-neighbor} (K-NN) classifier as a strong baseline and then propose TITAN (Tcr epITope bimodal Attention Networks), a bimodal neural network that explicitly encodes both TCR sequences and epitopes to enable the independent study of generalization capabilities to unseen TCRs and/or epitopes.
 \\
\textbf{Results:}
By encoding epitopes at the atomic level with SMILES sequences, we leverage transfer learning \replaced{and data augmentation}{techniques} to enrich the input data space and boost performance. TITAN achieves high performance \added{in the prediction of specificity of} unseen TCRs (ROC-AUC 0.87 in 10-fold CV) and surpasses the results of the current state-of-the-art (ImRex) by a large margin. 
\added{Notably, our Levenshtein-distance-based K-NN classifier also exhibits competitive performance on unseen TCRs.} While \added{the generalization to }unseen epitopes remains challenging, we report two major breakthroughs. First, by dissecting the attention heatmaps, we demonstrate that the sparsity of available epitope data favors an implicit treatment of epitopes as classes. This may be a general problem that limits unseen epitope performance for sufficiently complex models. Second, we show that TITAN nevertheless exhibits significantly improved performance on unseen epitopes and is capable of focusing attention on chemically meaningful molecular structures.
\\
\textbf{Availability:} 
\replaced{
The code as well as the dataset used in this study is publicly available at~\href{https://github.com/PaccMann/TITAN}{https://github.com/PaccMann/TITAN}.
}{
All code will be made available publicly on \texttt{GitHub} upon acceptance of our manuscript.
}\\
\textbf{Contact:} \href{wbr@zurich.ibm.com}{wbr@zurich.ibm.com}\\
\textbf{Supplementary information:} Supplementary data will be available at \textit{Bioinformatics}
online.
}

\maketitle

\section{Introduction}

T-cells are an integral part of the adaptive immune system, whose 
survival, proliferation, activation and function are all 
governed by the interaction of their T-cell receptor (TCR) with immunogenic peptides (epitopes) presented on major histocompatibility complex molecules (MHC).
A large repertoire of T-cell receptors with different specificity is needed to provide protection against a wide range of pathogens. This repertoire is generated using stochastic gene recombination and can theoretically produce diversities of $10^{15}$ -- $10^{20}$ different receptors~\citep{Laydon2015} in an individual, each with unique binding capabilities.
Due to this diversity, reliably predicting the binding specificity of a TCR from its sequence and understanding the mechanisms underlying TCR--pMHC interaction is highly challenging.
At the same time, it has enormous potential to transform the field of immunology.
A reliable prediction tool could unlock the wealth of information encoded in a patients' TCR repertoire, which reflects their immune history and could inform about past and current infectious diseases, vaccine effectiveness or autoimmune reactions. 
Additionally, it could empower the application of therapeutic T-cells for cancer treatment, allowing the study of effectiveness and potential cross-reactivity risks \textit{in silico}.

Recent advances in high-throughput sequencing techniques have led to the generation of \added{an increasing amount of }datasets linking TCR sequences to the epitopes they bind. However, the available data is still extremely sparse compared to the high dimensionality of the search space created by the TCR \added{theoretical }diversity. Moreover, current experimental settings typically link many TCRs to a single epitope, which leads to datasets that contain information about tens of thousands of TCRs, but only a few hundred different epitopes. 

Nevertheless, several studies have attempted the prediction of TCR specificity from sequence using machine learning (for a review see~\citet{Mosch2019a}).
The most successful approaches so far are categorical epitope models, which exploit the relative abundance of TCR sequences to learn patterns of TCRs binding to the same epitope. Various machine learning concepts were applied to this task, including decision trees~\citep{DeNeuter2018, gielis2019}, \deleted{and }a range of different clustering approaches~\citep{Glanville2017, Dash2017, Jokinen2019}\added{ and Variational Autoencoders~\citep{sidhom2021deeptcr}}. Many of these can successfully predict which one of a small set of epitopes a given TCR will most likely bind to. 
However, these approaches are inherently incapable of predicting specificity to epitopes not contained in the training set (unseen epitopes), which fundamentally limits their applicability. 

The next milestone \added{ towards this challenging goal} are generic models, which explicitly encode \added{both the TCRs and the }epitopes. These have the potential to predict binding of any TCR--epitope pair, \added{opening the door to the development of models that can generalize} to both, unseen TCRs and epitopes. 
Current models show moderate performance on test data containing epitopes already encountered in training, but cannot extrapolate to unseen epitopes~\citep{Jurtz2018, Springer2019, moris2020current}.  

TITAN \added{(Tcr epITope bimodal Attention Networks) exploits} a bimodal neural network \added{architecture to} explicitly encode both TCR and epitope sequences. \added{More specifically, TITAN }uses convolutions to aggregate local information and fuses the modalities, using an interpretable attention mechanism \added{from which binding probabilities are predicted.}

Since interpretability is paramount in healthcare applications of machine learning, 
the use of context attention is central to our model, as it allows us to explain the choices of the algorithm and to analyze which amino acids or even atoms the model focuses on.
These highlighted entities can be interpreted in the context of the underlying biochemical processes.
We explore different encoding strategies \added{for the epitopes} such as SMILES~\citep{Weininger1989}, a string-based, atom-level representation of molecules. SMILES are ubiquitously utilized in chemoinformatics \added{for a wide range of applications, from predicting the }chemical or pharmacological properties of molecules~\citep{Goh2017,manica2019paccmann} to generative modelling~\citep{Gomez-Bombarelli2018}. Using SMILES effectively results in a re-formulation of the TCR--epitope binding problem as the more general compound protein interaction (CPI) task, thus enabling the usage of large databases of protein-ligand binding affinity \added{for pretraining, }e.g. BindingDB \added{including} >1M labelled samples~\citep{gilson2016bindingdb}. 
\section{Methods}
\vspace{-2mm}
\subsection{Data}\label{sec:data}
In order to assemble a \added{larger and more diverse} dataset, we combine data collected in the VDJ database~\citep{Bagaev2019} with a recently published COVID-19 specific dataset published by the ImmuneRACE project~\citep{immunecode2020}. 
Since paired chain data is still rare, we restrict ourselves to TCR$\beta$ chain sequences.

We use all human TCR$\beta$ sequences downloaded from the VDJ database (07/12/2020), i.e. 40,438 TCR sequences assigned to 191 peptides.
Since this dataset is highly imbalanced, we exclude epitopes with less than 15 associated TCR sequences and downsample to a limit of 400 TCRs per epitope\deleted{ to build a more balanced dataset}. After these preprocessing steps, we are left with a dataset of 10,599 examples on 87 epitopes. We refer to this dataset as \textit{VDJ}.

The COVID-19 dataset (published 25/07/2020) originally contained 154,320 examples associated with 269 different epitopes or groups of epitopes. 
To avoid ambiguity, we keep only samples associated with a single unique epitope and exclude unproductive sequences. 
Then we apply the same preprocessing steps as for the VDJ dataset, downsampling to 400 TCRs/epitope and excluding epitopes with less than 15 associated TCRs to arrive at a dataset of 12,996 examples. 

We refer to the combined dataset with samples from VDJ and the COVID-19 dataset as \textit{VDJ+COVID-19}.

Since these primary datasets contain only positive examples, we need to generate negative data. 
This can be achieved by shuffling the sequences, thereby associating TCRs with epitopes that they have not been shown to bind. 
Due to the low probability of a randomly drawn TCR binding a specific epitope, this manner of generating negative samples is established in the field~\citep{moris2020current, Fischer2020}. It has also been shown to limit overestimation of performances in comparison to adding additional naive TCR sequences from other sources~\citep{moris2020current}.
\replaced{Furthermore, by shuffling the pairing of TCRs and epitopes, we can match the number of negative examples to that of positive examples}{generate the same number of negative examples as positive examples} for each TCR, \added{avoiding unbalanced datasets. With this procedure}, we build a training dataset of 46,290 examples, 50 \% of which are positive, encompassing 192 different epitopes.

To ensure a fair comparison to the state of the art model ImRex, we also downloaded the \added{publicly available} dataset that ImRex was trained on\deleted{ , which was made publicly available}. \added{This dataset is based on the VDJ database and contains 13,404 samples for 118 different epitopes, with 50 \% negative samples. }
We use it to train all models for the final comparison (see section \ref{ssec:imrex}). To evaluate the performances, we use an independent test set generated from the McPAS database~\citep{tickotsky2017mcpas} (downloaded on 03/11/2020). We exclude non-human TCRs and remove all samples with TCRs contained in the ImRex training data. Then we split the McPas data into two test sets: one \replaced{including}{containing} all samples with epitopes contained in the ImRex training set (seen epitope test set) and one with epitopes not contained in the ImRex dataset (unseen epitope test set). For both test sets, 50\% negative data is generated by shuffling. The final seen epitope test set contains 9740 samples, the unseen epitopes test set contains 1458 samples.

\subsection{Models}
\begin{figure*}[ht!]
	\centering
	\includegraphics[width=\textwidth]{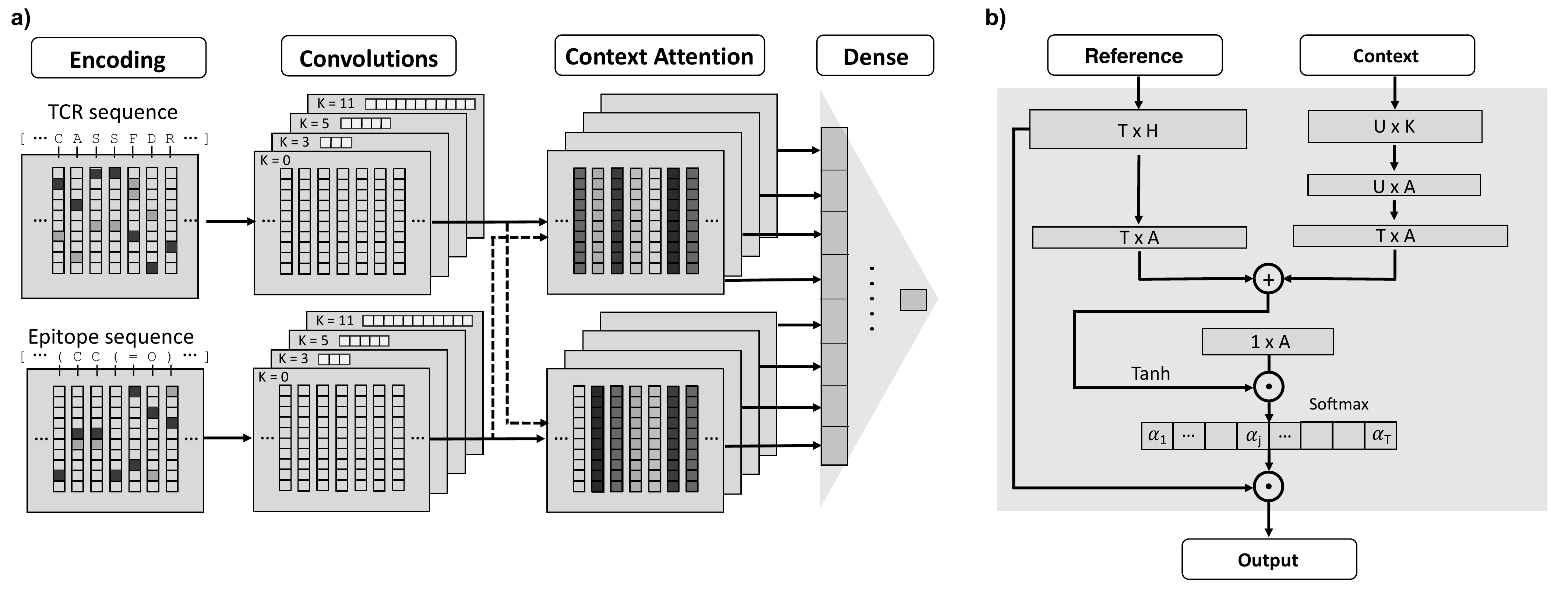}
	\vspace{-5mm}
	\caption{\textbf{Overview of TITAN architecture.} \textbf{a)} Our model ingests a TCR and an epitope sequence, which get encoded using BLOSUM62 for amino acid sequences or learned embeddings for SMILES.
	Then, 1D convolutions of varying kernel sizes are performed on both input streams before context attention layers generate attention weights for each amino acid of the TCR sequence \textit{given an epitope} and vice versa. 
	Finally, a stack of dense layers outputs the binding probability. 
	\added{Conceptually, this architecture is identical to the one proposed in~\citet{born2021datadriven} (cf. Figure S3) but our visualization here is more fine-grained.
	}
	\textbf{b)} The linchpin of the model is the bimodal context attention mechanism.
	It ingests the convolved TCR and epitope encodings, treating one as \textit{reference} and the binding partner as \textit{context}.
	A series of transformations combines the modalities and yields an attention vector over the reference sequence (driven by the context) that can be overlayed with the molecule like a heatmap.
	}
	\label{fig:ML}
\end{figure*}
\label{ssec:models}

\subsubsection{Problem formulation}
\label{sssec:problem}
We are interested \added{in learning a mapping $\Phi$ between the space of receptors $\mathcal{T}$ and the space of epitopes $\mathcal{E}$ to the the space of affinity scores $\mathcal{A}$, i.e}
$\Phi: \mathcal{E}\times \mathcal{T}\rightarrow \mathcal{A}$.
$\Phi$ is learned with a training data set $\mathcal{D}=\{e_i, t_i, a_i\}_{i=1}^N$ where $e_i \in \mathcal{E},t_i \in \mathcal{T}$ and $a_i\in \{0,1\}$ is a binary label \replaced{indicating}{representing} whether binding occurred.

\subsubsection{K-NN Baseline}
\label{sssec:knn}
Our baseline model for the presented TCR-epitope binding prediction is constituted by a $k$-nearest-neighbor (K-NN) classifier.
As a distance metric between samples we utilize the sum of the length-normalized Levenshtein distance of the respective epitope and TCR protein primary sequences. \\
More formally, for the training data $\mathcal{D}=\{e_i, t_i, a_i\}_{i=1}^N$, we choose $e_i$ and $t_i$ to be \added{epitope and }\replaced{TCR}{protein} sequences, respectively ($t_i$ is the CDR3 region).
Let $\{e_j, t_j\}$ denote an unseen sample from the test dataset  $\mathcal{D}_{Test}=\{e_i, t_i\}_{i=1}^{N_{Test}}$.
With the goal of predicting $\hat{a}_j$ to approximate the unknown $a_j$, we first retrieve the subset of training data $\mathcal{D}_k$ containing the $k$ nearest neighbors using the distance measure
\begin{equation}
\label{eq:knn}
    \mathbf{D}(e_i, t_i, e_j, t_j) =\frac{Lev(e_i, e_j)}{|e_j|} + \frac{Lev(t_i, t_j)}{|t_j|}
\end{equation}
where $|\cdot|$ denotes sequence length and $Lev(\cdot,\cdot)$ is the Levenshtein distance~\citep{levenshtein1966binary}, i.e., a string-based distance measure that measures the number of single-AA changes required to \replaced{transform}{morph} one sequence into the other. 
Then, the prediction $\hat{a}_j$ is trivially computed by $\hat{a}_j = \frac{\sum_i^{k}a_i}{k}$ with $a_i \in \mathcal{D}_k$. 
\deleted{In practice,} We evaluate the model on all odd $k$\added{ (to avoid ties)}, s.t. $1\leq k \leq 25$, \added{and choose the value for $k$ leading to the best ROC-AUC score for comparisons.}.
Note that this model is non-parametric. \deleted{, s.t. $\Phi=\emptyset$.}

\subsection{Model Architecture}
Fig.~\ref{fig:ML} shows an overview of the algorithmic steps of TITAN. 
To predict binding, we devise a bimodal architecture that separately ingests both the TCR and the peptide sequence.

The TCR sequences are encoded with the BLOSUM62 matrix~\citep{Henikoff1992}, which is based on evolutionary similarity of amino acids and has been widely applied in TCR specificity prediction tools~\citep{Jurtz2018, Jokinen2019}.  Either the full sequence or only the CDR3 region were used as input.
Since the antigenic peptides are small molecules, we explore two options to encode them\added{: one that uses} an amino acid-wise encoding such as BLOSUM62, and \deleted{alternatively,} \added{ one that uses} an atom-level encoding with SMILES. 
All sequences were padded to the same length of 500 tokens.
Advantageously, SMILES representations of a molecule are not unique, thus facilitating data augmentation~\citep{Bjerrum2017}.

The remaining architecture is inspired by~\citet{manica2019paccmann} and almost identical to the compound-protein-interaction (CPI) model presented in~\citet{born2021datadriven}.
\added{
In case of pretraining on CPI data (see below), the models are identical, otherwise the SMILES-encoding ligand input stream is replaced with an AA-encoding epitope stream.
}
Three parallel channels with convolutions of kernel sizes 3, 5 and 11 are employed on the input sequences to combine information from local neighbourhoods of varying spatial extend.
A fourth channel has a residual connection without convolutions (see Fig.~\ref{fig:ML}a).
For each of the four channels, we utilize two attention layers, where one modality 
is used as a context to compute attention scores over the 
other
(see Fig.~\ref{fig:ML}b).
This allows the model to use information from the binding partner\added{, i.e the context, to} learn the importance of each token in the input sequence, i.e. the reference.
The attention weights $\alpha_i$ are computed as: 
\vspace{-1mm}
\begin{equation}
\vspace{-1mm}
    \alpha_i = \frac{\exp{(u_i)}}{\sum_{j}^{T}{\exp{(u_j)}}} \quad \text{, where} \quad \vec{\textbf{u}} = \tanh{ \left( \mathbf{X}_{1}\mathbf{W}_{1} + \mathbf{W}_{3}(\mathbf{X}_{2}\mathbf{W}_{2})\right) \vec{\mathbf{v}}}
\end{equation}
We call $\mathbf{X}_{1}\in\mathbb{R}^{T\times H}$ the \textit{reference} input, where $T$ is the sequence length and $H$ is the number of convolutional filters.
Further, $\mathbf{X}_{2}\in\mathbb{R}^{U\times K}$ is the \textit{context} input, where $U$ and $K$ are sequence length and number of convolutional filters in the other modality, respectively.
$\mathbf{W}_{1}\in\mathbb{R}^{H\times A}$, $\mathbf{W}_{2}\in\mathbb{R}^{K\times A}$, $\mathbf{W}_{3}\in\mathbb{R}^{T\times U}$ and $\vec{\mathbf{v}}\in\mathbb{R}^{A}$ are learnable parameters.
Intuitively, both inputs are projected into a common attention space $\mathbb{R}^{A}$ with $A = 16$ and then summed up, which enables the layer to take the context into account for determining feature relevance. 
$\vec{\mathbf{v}}$ combines the information through a dot product, the output of which is fed to a softmax layer to obtain the attention weights $\alpha_i$, which are used to filter the inputs.
Finally, both TCR and peptide information gets passed to two dense layers with 368 and 184 nodes, respectively, which output the binding probability. 

\subsection{Pretraining}
By using SMILES encodings of the epitopes, predicting epitope receptor binding affinity can be seen as a CPI prediction task.
We utilize BindingDB~\citep{gilson2016bindingdb}, as of April 2020, to pretrain our model. 
To reduce \added{the problem complexity and potential biases associated with the different experimental platforms to measure affinity, we binarize the binding data and  }define all entries in the database as binding, ignoring \added{continuous} affinity measurements. We generate an equal amount of negative examples by randomly assigning ligands to proteins~\citep{born2021datadriven}.
In order to avoid high discrepancy between sequence lengths, ligands with a length > 250 SMILES tokens and proteins larger than 1028 amino acids are discarded.
This results in 325,688 ligands and 3351 proteins and a total of 471,017 pairs from which 90\% (423,915) are used for training and the rest for validation.
\added{To adapt the model size to the larger available dataset, we changed the layer sizes for pretraining by padding TCR sequences to 1028, setting the attention space $A=256$, using four convolutional channels with kernel sizes 3, 7, 9 and 13 for epitopes and 3, 7, 13, 19 for TCR and using three final dense layers with 2048, 1024 and 512 nodes.}
\begin{figure}[b]
	\centering
	\includegraphics[width=0.45\textwidth]{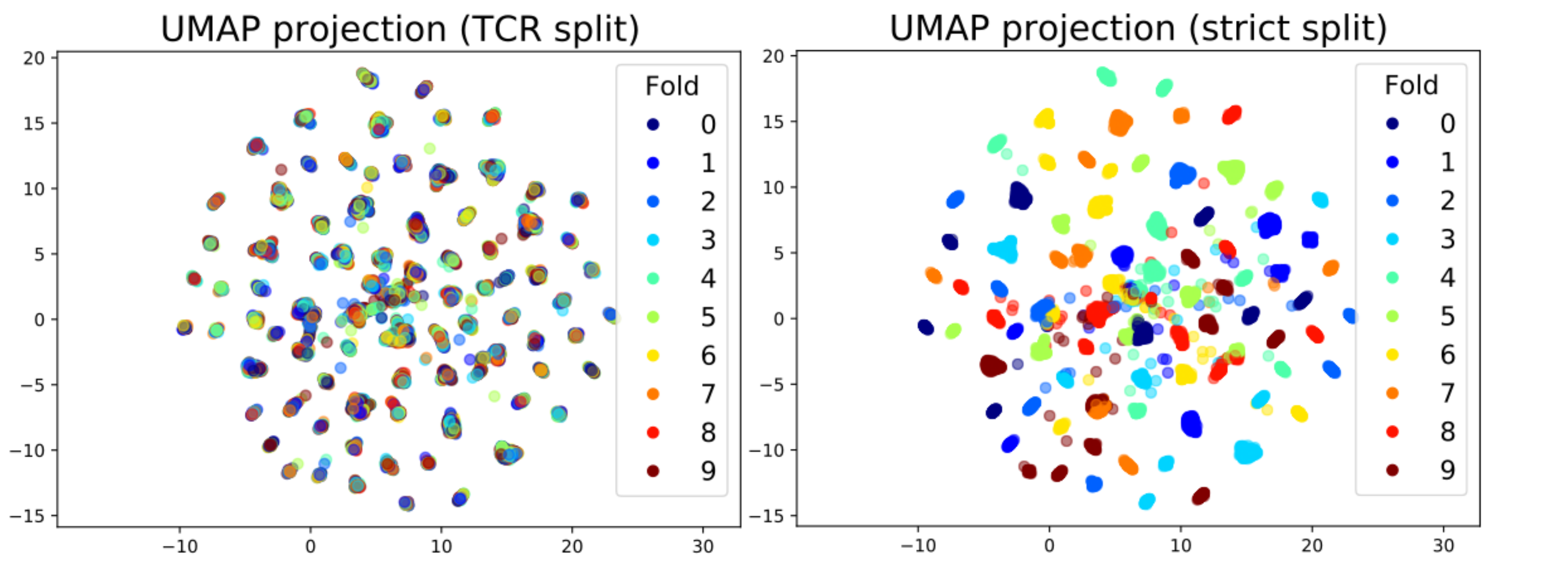}
	\caption{
	\textbf{UMap Visualization of the VDJ dataset colored by \replaced{the fold each sample belongs to}{class belonging} in the TCR split (left) and strict split (right).}
	\replaced{The UMap projection leads to clear clusters of the samples. Coloring by fold in the TCR split reveals to connection between clusters and folds. However, coloring by fold in the strict split reveals that all the samples in a cluster belong to the same fold. This suggests that clusters correspond to distinct epitopes, highlighting their heterogeneity and the}{No evident structure is visible in the TCR split, but the coloring in the strict split that clusters samples according to their epitope identity shows the heterogeneity of the data and difficulty} \added{challenge} of good generalization for the strict split.
	}
	\label{fig:umap}
\end{figure}
\subsection{Data Splitting}
We evaluate our models on a 10-fold cross-validation split. To determine the generalization capabilities of the models towards unseen TCRs and unseen epitopes separately, we use two different split methods. In the first, we ensure that each TCR is restricted to only one fold, which ensures that the validation datasets do not contain TCRs which were shown in training. The epitopes, however, are distributed randomly over the folds, so that most of the epitopes in the validation dataset were also shown during training.
We refer to this split as the \textit{TCR split}. 
Additionally, we generate a \textit{strict split}, where we ensure that each TCR and each epitope is restricted to a single fold, ensuring that neither TCRs nor epitopes contained in the validation dataset were shown during training. To ensure the separation of TCRs and epitopes in their folds, we generate negative data by shuffling within each fold.
A UMap~\citep{mcinnes2018umap} visualization of all samples of the dataset is presented in Fig.~\ref{fig:umap} and shows the ramifications of the splitting strategy. 
The feature space for the UMap dimensionality reduction was generated by embedding the amino acid sequences using a pretrained protein language model~\citep{elnaggar2020prottrans}.

\subsection{Model Training}
All described architectures were implemented in \texttt{PyTorch} 1.4\added{ and used the \texttt{pytoda} package for data handling and preprocessing. 
The models }optimized binary cross entropy loss with Adam~\citep{Kingma2015} ($\beta_1$ = 0.9, $\beta_2$ = 0.999, $\epsilon$ = 1e-8) and a learning rate of 0.0001.
In the convolutional and dense layers we employed dropout ($p$ = 0.5) and ReLU activation. All models were trained with a batch size of 512 on a cluster equipped with POWER8 processors and a single NVIDIA Tesla P100.
\added{
The learning rate was tuned using the VDJ dataset and the remaining hyperparameters were chosen carefully based on previous experience.
}

\FloatBarrier
\section{Results}
\subsection{Performance on TCR Split}

\begin{figure*}[ht!]
	\centering
	\includegraphics[width=0.9\textwidth]{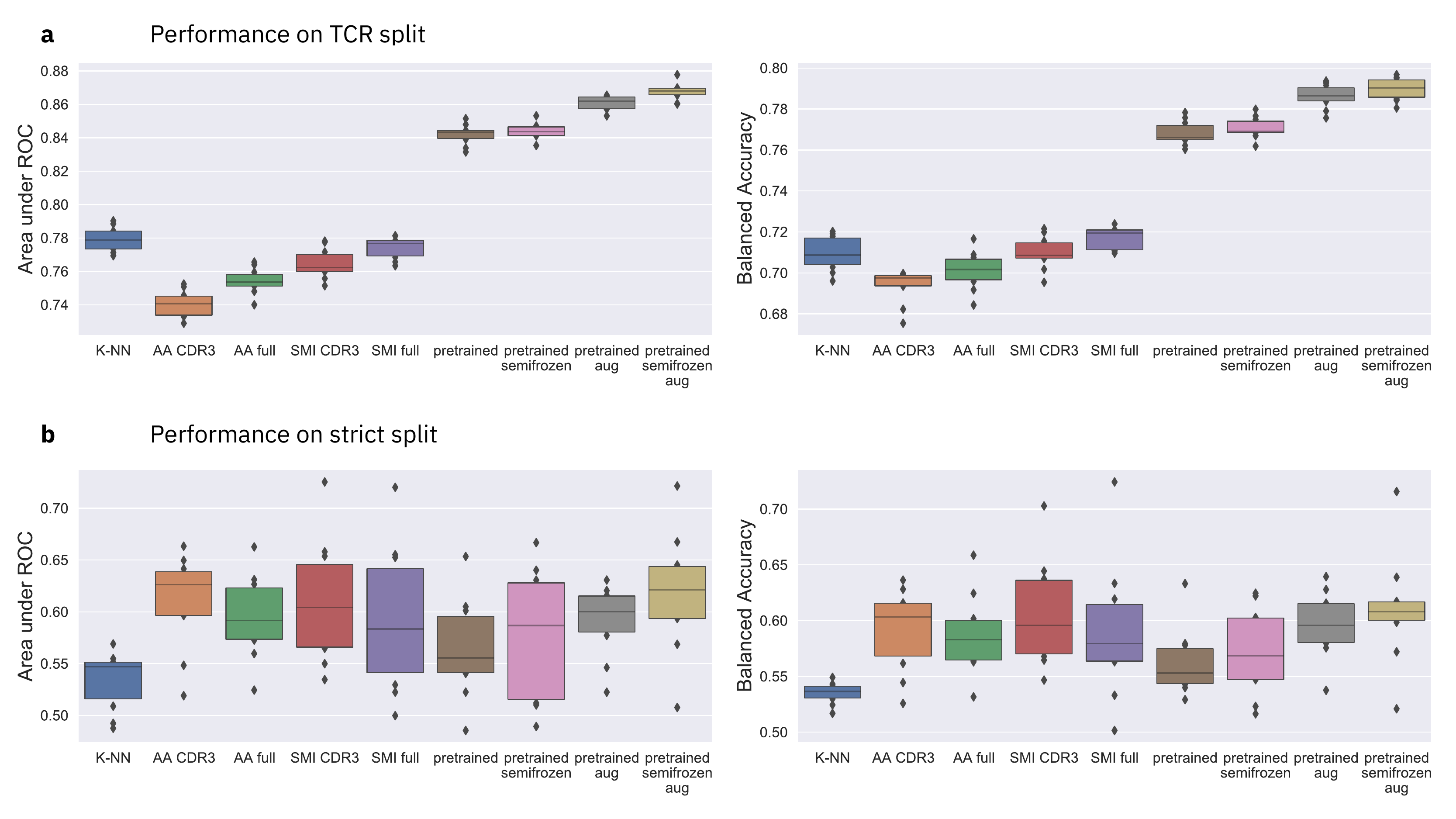}
	\vspace{-5mm}
	\caption{Performance comparison of different TITAN model settings \added{trained or fine-tuned on the VDJ+Covid dataset}. \textbf{a)} ROC-AUC scores and balanced accuracy on a 10-fold crossvalidation TCR split (validation and training data share epitopes, but not TCRs). \textbf{b)} ROC-AUC scores and balanced accuracy on a strict 10-fold crossvalidation split, where validation and training data share neither epitopes, nor TCRs. All boxplots: The center of each boxplot marks the sample median, and the box extends from lower to upper quartile. \textit{K-NN} refers to the baseline model. Other abbreviations denote different settings under which TITAN was trained.
	\textit{AA CDR3:} amino acid encoding of epitopes, only CDR3 sequence input for TCRs; \textit{AA full:} amino acid encoding of epitopes, full sequence input for TCRs;
	\textit{SMI CDR3:} SMILES encoding of epitopes, only CDR3 sequence input for TCRs;
	\textit{SMI full:} SMILES encoding of epitopes, full sequence input for TCRs;
	\textit{Pretrained:} SMILES encoding of epitopes, full sequence input for TCRs, model pretrained on BindingDB;
	\textit{Pretrained semifrozen:} SMILES encoding of epitopes, full sequence input for TCRs, model pretrained on BindingDB, weights in epitope channel fixed during fine-tuning;
	\textit{Pretrained aug:} SMILES encoding of epitopes, full sequence input for TCRs, model pretrained on BindingDB with data augmentation;
	\textit{Pretrained semifrozen aug:} SMILES encoding of epitopes, full sequence input for TCRs, model pretrained on BindingDB with data augmentation, weights in epitope channel fixed during fine-tuning.
	 }
	\label{fig:performance}
\end{figure*}

In Fig~\ref{fig:performance}a, we compare 10-fold crossvalidation performances \added{of different TITAN settings} on the TCR split scenario, which allows us to gauge the generalization capabilities of the algorithm towards unseen TCR sequences\deleted{accross different model settings}.
We \replaced{explored}{have} several options to input the sequence information of TCR and epitope in our model. 
\added{Following the well-established concept of word embeddings~\citep{mikolov2013efficient}, we can encode the amino acids using a fixed-length vector representation (in our case 32-dimensional), that is initialized randomly and then learned through training. 
Alternatively, we can represent each amino acid as a vector containing biophysical properties (e.g. molecular weight, residue weight, pKa, pKb, pKx, pI, and hydrophobicity at pH2). 
Finally, we explored using evolutionary substitution matrices like BLOSUM62. 
Each row in the BLOSUM matrix represents the probability for an amino acid to be substituted by any of the other amino acids and can be used as a 26-dimensional vector representation of that amino acid. 
}

\deleted{The simplest approach is to directly encode the amino acid sequence of both TCR and epitope, using either biophysical features, or the BLOSUM matrix, or a learned embedding.} 

An initial comparison of these embedding options showed no clear preference for either of them (\added{See Table}~\ref{tab:embedding_comparison}). Since a learned embedding is less reproducible and biophysical feature choices can be debated, we decided to use the BLOSUM62 matrix to embed amino acid sequences in all model settings. 

\begin{table}[h]
    \centering
    \caption{\textbf{Comparison of Amino Acid Embeddings.}
    Mean and standard deviation of the \textit{AA CDR3} model configuration on 10-fold TCR split \added{of the VDJ dataset.}\added{ All tested amino acid embeddings \deleted{choices} show a similar performance. }
    }
    	\begin{tabular}{cc}
    	    \textbf{Embedding}  & \textbf{ROC-AUC} \\
    	    \toprule
    	    Biophysical Features & $0.76 \pm 0.01$\\
    	    \hline
    	    Learned Embedding & $0.75 \pm 0.01$\\
    	    \hline
    	    BLOSUM62 Matrix & $0.75 \pm 0.01$\\
    	    \hline
    	\end{tabular}
    	\label{tab:embedding_comparison}
\end{table}

For TCRs, we have the options to either focus solely on the hypervariable CDR3 loop, which is known to be the main peptide binding region, or to consider the full variable region of the TCR$\beta$ sequence, which includes the V, D and J segments and encompasses all three hypervariable loops CDR1, CDR2 and CDR3. Fig~\ref{fig:performance}a shows that the use of the full sequence information boosts the model performance, indicating that valuable information is contained in regions outside of the CDR3 loop. 

Since epitopes are relatively short (5 to 15 amino acids), we can represent them in a fine-grained, atom-wise manner using SMILES strings. We can see in Fig~\ref{fig:performance}a that the SMILES representation of epitopes further improves performance compared to an amino acid encoding. 

The combination of SMILES for epitopes and full sequence encoding of the TCR leads to a mean ROC-AUC of $0.774 \pm 0.006$ and a mean balanced accuracy of $0.716 \pm 0.005$. However, we see that even this improved model does not outperform the simple K-NN baseline model that we included as a comparison. \added{With $k = 13$, the K-NN baseline achieves the best results with} a ROC-AUC of $0.779 \pm 0.007$ and a balanced accuracy of $0.709 \pm 0.008$ \added{(see Fig.~S1)}. This highlights the importance of including an \added{appropriate} baseline model, which is so far rarely observed in the field, as simple models may outperform complex ones in a sparse data setting. 
We emphasize that this is not an argument \textit{against} our neural networks, but \textit{for} the K-NN baseline model, which also outperforms the \added{state of the art model,} ImRex, a recent approach that uses 2D CNNs~\citep{moris2020current} (see section \ref{ssec:imrex}\added{ for a more detailed comparison \replaced{of}{between} TITAN and ImRex).}
\added{We also note that} an approach similar to our K-NN baseline\added{, TCRMatch~\citep{chronister2020tcrmatch},} was recently presented in a preprint. TCRMatch \added{predicts} TCR specificity \added{using only}  sequence similarities to previously characterized receptors. 

All model comparisons on the TCR split are also summarized in Table~\ref{tab:tcr_split_rocauc}. For better comparability to previously published models, we also include scores obtained on the dataset excluding the samples gathered from the COVID-19 dataset.

\begin{table}[b]
    \caption{\textbf{10-fold cross validation performance on TCR split.}
    Mean and standard deviation of each model configuration on the VDJ dataset and the VDJ + COVID-19 dataset.
    }
    	\begin{tabular}{ccc}
    	    & \multicolumn{2}{c}{\textbf{ROC-AUC}}\\
    	    \textbf{Model}  & \textbf{VDJ} & \textbf{VDJ+COVID-19}\\
    	    \toprule
    	    K-NN (Baseline) &
    	    $0.789\pm0.01$  & $0.78\pm0.01$\\
    	    \hline
    	    AA CDR3 & 
    	    $0.75 \pm 0.02$&
    	    $0.74 \pm 0.007$	\\
    	    \hline
    	    AA full &
    	    $0.76 \pm 0.007$ &
    	    $0.75 \pm 0.007$ \\
    	    \hline
    	    SMI CDR3 &
    	    $0.73 \pm 0.007$ &
    	    $0.76 \pm 0.008$\\
    	    \hline
    	    SMI full &
    	    $0.75 \pm 0.006$&
    	    $0.77 \pm 0.006$\\
    	    \hline
    	    Pretrained & 
    	    $0.81\pm 0.01$ &  $0.841\pm 0.01$ \\
    	    \hline
    	    Pretrained aug. &
    	    $0.803\pm 0.01$ & $0.861\pm 0.00$ \\
    	    \hline
    	    Pretrained semifrozen &
    	    $0.801\pm 0.01$ & $0.844\pm 0.00$ \\
    	    \hline
    	    Pretrained semifrozen aug. & 
    	    $\mathbf{0.822\pm 0.01}$ & $\mathbf{0.868\pm 0.01}$\\
    	    \hline
    	\end{tabular}
    	\label{tab:tcr_split_rocauc}
\end{table}

\added{Furthermore, the  results in Table~\ref{tab:tcr_split_rocauc} also }indicate that the available interaction data may be too sparse to enable the models to learn the complex interaction patterns governing TCR--epitope binding. However, using the SMILES encoding for epitopes and the full sequence encoding for TCRs, we have effectively re-formulated the task as a compound protein interaction, which  allows us to use BindingDB~\citep{gilson2016bindingdb} to pretrain the model, before we fine-tune it on the TCR--epitope interaction data. 
The pretrained model performed well on the held-out data from BindingDB (ROC-AUC 0.895).

\added{Regarding the pretraining,} we tested two different settings, one where all weights could be adapted during fine-tuning, and a \textit{semifrozen} setting, where we only allowed weight changes in the TCR channel and the final dense layers of the epitope. 
This was done to prevent \replaced{catastrophic interference of}{the model from "unlearning" to recognize relevant} SMILES features due to the extremely low number of different epitopes in the fine-tuning dataset. Fig~\ref{fig:performance} shows that pretraining severely improves model performance in both settings. We further improved model performance by exploiting the non-uniqueness of SMILES strings to perform data augmentation. \added{Augmentation is achieved by randomly generating a valid SMILES representation~\citep{Bjerrum2017} of the epitope on the fly at each training step using \texttt{pytoda}~\citep{born2021paccmannrl}}. The best overall model performance was achieved by the semifrozen pretrained model with augmentation, with a mean ROC-AUC of $0.868 \pm 0.005$ and a mean balanced accuracy of $0.790 \pm 0.005$, clearly outperforming the K-NN baseline by a large margin.

\begin{figure*}[ht!]
	\centering
	\includegraphics[width=0.9\textwidth]{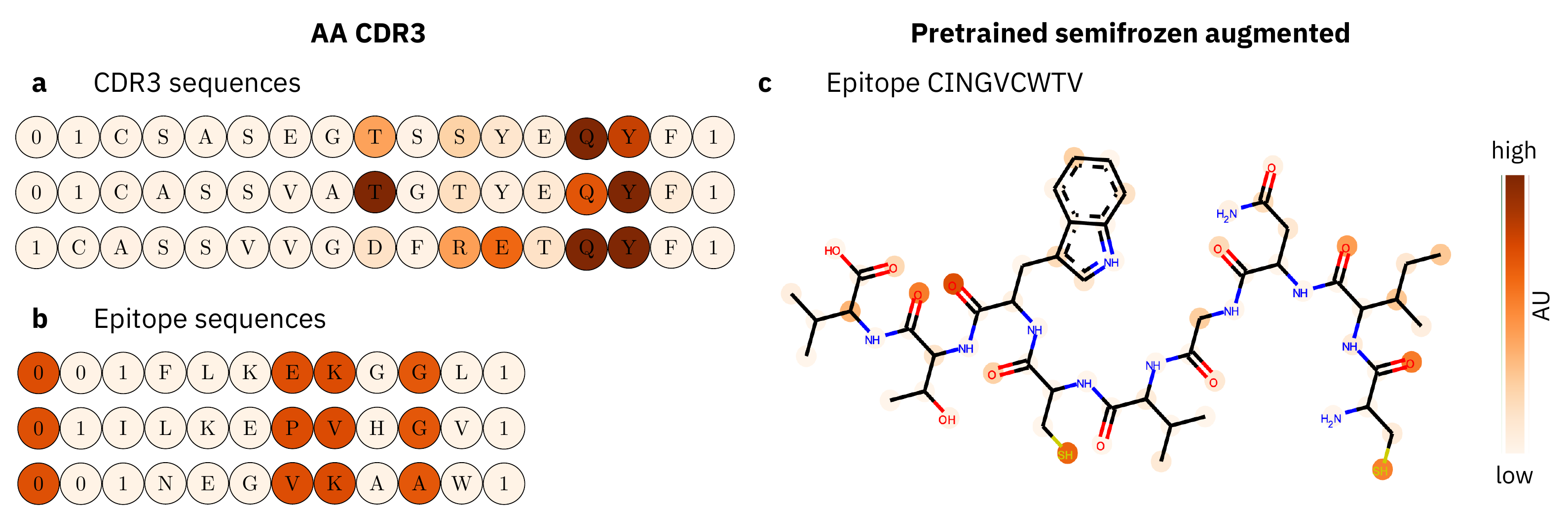}
	\vspace{-5mm}
	\caption{Analysis of the context attention layers. The attention weights $\alpha_i$ on each token are extracted from the context attention layer and represented as a colormap. In \textbf{a)} and \textbf{b)}, examples of the \textit{AA CDR3} setting are shown, where epitopes are input as amino acid sequences and only CDR3 sequences are input in the TCR \added{channel}. $0$ denotes the padding and $1$ the \texttt{<START>} and \texttt{<STOP>} token. \textbf{a}) shows the attention scores over three exemplary CDR3 sequences. The model adapts the attention scores based on the different inputs. \textbf{b}) shows the attention on three exemplary epitopes. The model fails to adapt the attention scores to different inputs. 
	\textbf{c)} Attention scores over an exemplary epitope that was not included in the training dataset. The model was pretrained on BindingDB and fine-tuned on the training data with a frozen epitope input channel with SMILES augmentations to enrich the data. 
	The lower bound of the color scale was set to one standard deviation above the mean attention on the padding tokens. 
	}
	\label{fig:attention}
\end{figure*} 

The high performance on validation data that does not contain TCR sequences used in training shows that the model successfully generalizes to unseen TCRs. 
\added{Comparing the performance across groups of TCRs with different similarities to the training data, we find that while the model performs better for TCRs that are highly similar to their closest partners in the training set, the performance on the TCRs with highest distance from the training data is still high (ROC AUC $0.844 \pm 0.016$, balanced accuracy $0.709 \pm 0.008$, see Fig~S3).}

\subsection{Analysis of Attention Layers}
\label{ssec:attention_analysis}

We can investigate the decision processes of TITAN using the information contained in the attention layers.
Fig.~\ref{fig:attention}a shows the attention scores of TITAN in the \textit{AA CDR3} setting \added{-- a setting that clearly outperforms  previous work~(see subsection~\ref{ssec:imrex}) -- }over a number of \added{exemplary} CDR3 sequences.
We see that while there is some preference to focus on certain positions, the model adapts the attention to the different sequences. The mean inter-TCR variance per token is at $4.3 \cdot 10^{-5}$. 
Moreover, we find that the attention also adapts to the context \added{(i.e., the epitope), for which we want to} predict the interaction. The mean variance per token within the same TCR interacting with different epitopes is $1.3 \cdot 10^{-5}$. This demonstrates that the model is capable of adapting the attention layer to both the input and the context. 

Fig.~\ref{fig:attention}b shows the attention of the same model \added{(\textit{AA CDR3} setting)} on several exemplary epitopes. 
We can clearly see that the model chooses to focus heavily on the same positions on each epitope. The \added{preferred} positions are independent of the sequence of both the input epitope and the context TCR. Comparing the attention scores across epitopes, we find that both the inter-epitope and the intra-epitope variance of the attention are extremely small, at $4.9 \cdot 10^{-10}$ and $2.5 \cdot 10^{-9}$ respectively.
This behavior indicates that TITAN fails to learn meaningful patterns in the epitope sequences from the limited diversity of epitopes in the dataset. 
We conjectured that the model finds a way to internally generate classes of epitopes represented by meaningless --- but unique --- vectors and predict specificity of TCRs towards these. 
This hypothesis is supported by the observation that compressing each epitope sequence to the chain of amino acids with attention scores $\alpha_i>0.1$ yields only a very moderate reduction in the number of unique sequences.
Crucially, these shortened epitope sequences are still unique for \replaced{185 out of 192}{225 out of 244} epitopes in the dataset \replaced{($96\%$)}{($92\%$)}.
By focusing on these fixed, invariant positions the model circumnavigates \replaced{learning}{to learn} generic representations of epitopes and instead internally classifies epitopes, at the cost of losing the power to differentiate 19 epitopes.

Pretraining the models on the compound interaction task greatly increased the diversity of sequences that were seen by the epitope channel of the model. While most of the ligands in BindingDB are not peptides, and  may therefore exhibit structures and chemical properties that differ strongly from the epitopes, the model may nevertheless infer general rules of chemical interaction from them. \added{This enables the model to learn more meaningful attention weights, which may be the reason for the performance boost we observe for pretrained models.}
In Fig.~\ref{fig:attention}b, we show a visual representation of the attention scores over one exemplary epitope as a case study of TITAN in the \textit{pretrained semifrozen augmented} setting. The chosen epitope \texttt{CINGVCWTV} was not shown during training.
The attention scores therefore reflect the transferable knowledge the model has gained during pretraining\deleted{ and fine-tuning}. We see that the attention patterns align well with our expectations. The attention is high on many of the oxygens, as well as on the two thiol groups of the cysteines. Nevertheless, we see that while the attention layers may extract some chemically relevant structures, they fail to capture others. Specifically, large and hydrophobic amino acid sidechains tend to get low attention scores, although they may be of high importance for molecule interactions.

\subsection{Performance on Strict Split}
Fig.~\ref{fig:performance}b compares the performances of the different TITAN settings on the strict epitope split, which measures the generalization capabilities to unseen epitopes. As expected, model performance drops severely across all settings. Moreover, we find that all TITAN settings perform similarly, with mean ROC-AUC scores around 0.6, while the K-NN baseline model shows a mean ROC-AUC of only $0.53 \pm 0.03$ \added{for a choice of $k=25$ (see Fig.~S2 for comparison)}. Compared to the unseen TCR performance, we also see an increase in the standard deviation of the scores, with ROC-AUC scores of over 0.7 for some folds, and 0.5 for others. All mean ROC-AUC values are summarized in Table~\ref{tab:strict_split_rocauc}. 

The best score overall is still achieved by the \textit{pretrained semifrozen augmented} model, with a mean ROC-AUC of $0.62 \pm 0.05$ and a mean balanced accuracy of $0.61 \pm 0.05$. However, its \added{superiority over other TITAN setting is not as large as in the TCR split scheme}. We can also see that pretraining does not strongly improve model performance on unseen epitopes. 

\begin{table}[h]
    \centering
    \caption{\textbf{10-fold cross validation performance on strict split.}
    Mean and standard deviation of each model configuration on the VDJ + COVID-19 dataset.
    }
    	\begin{tabular}{cc}
    	    \textbf{Model}& \textbf{ROC-AUC}\\
    	    \toprule
    	    K-NN (Baseline) &
    	    $0.535\pm0.03$\\
    	    \hline
    	    AA CDR3 & 
    	    $0.60 \pm 0.04$\\
    	    \hline
    	    AA full &
    	    $0.59 \pm 0.04$ \\
    	    \hline
    	    SMI CDR3 &
    	    $0.60 \pm 0.06$\\
    	    \hline
    	    SMI full &
    	    $0.59 \pm 0.06$\\
    	    \hline
    	    Pretrained & 
    	    $0.564\pm 0.04$ \\
    	    \hline
    	    Pretrained + Aug. &
    	    $0.591\pm 0.03$ \\
    	    \hline
    	    Pretrained semifrozen &
    	    $0.577\pm 0.06$ \\
    	    \hline
    	    Pretrained semifrozen + Aug. & 
    	    $\mathbf{0.618\pm 0.06}$\\
    	    \hline
    	\end{tabular}
    	\label{tab:strict_split_rocauc}
\end{table}

\begin{figure}[h]
	\centering
	\includegraphics[width=0.45\textwidth]{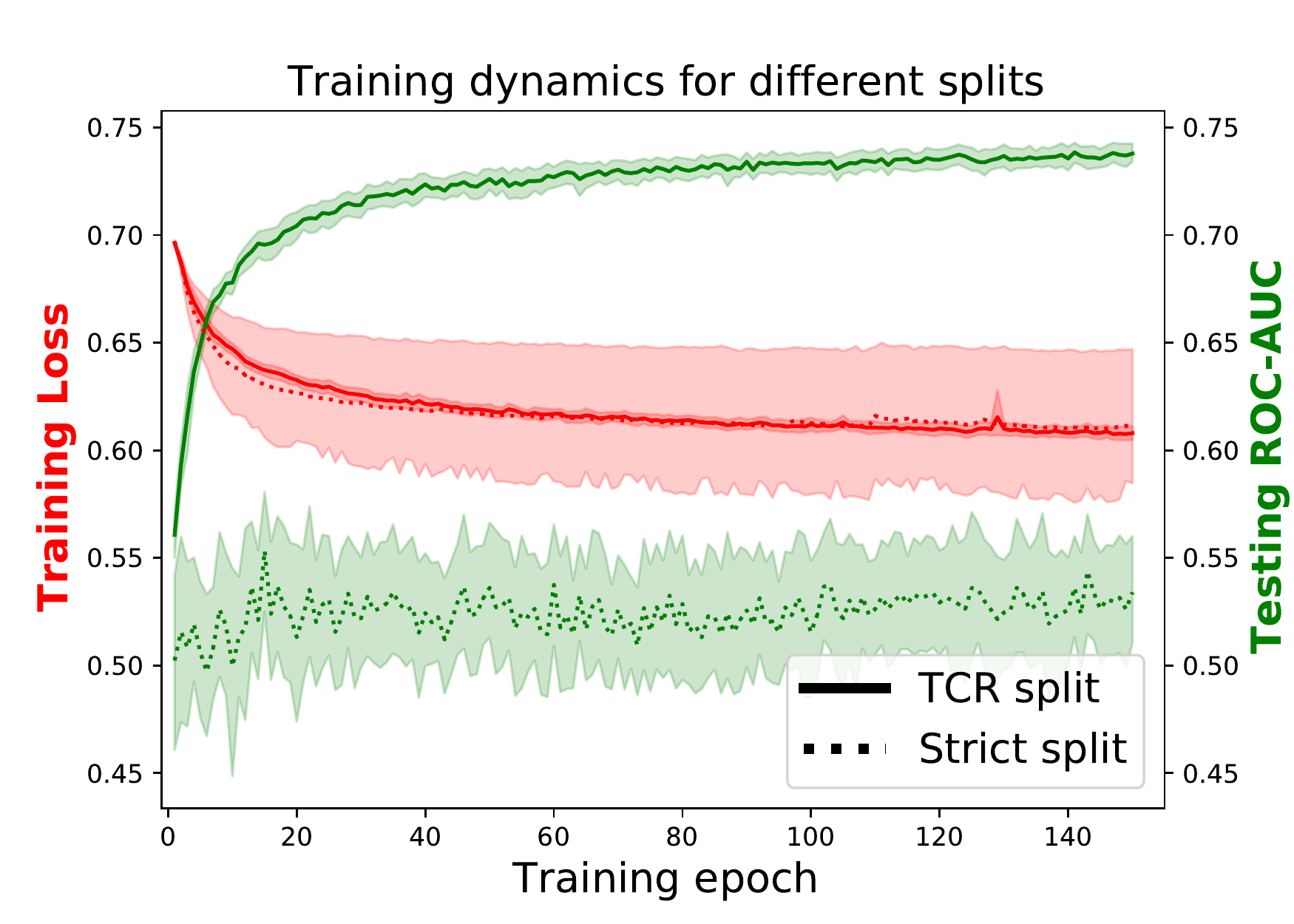}
	\caption{
	\textbf{Training dynamics for both splitting strategies.}
	A key distinction between the training on the TCR and strict splits is that the validation performance steadily converges only in the TCR split case.\added{ In the strict split case, no significant improvement is achieved even after training for more than 140 epochs.}
	This indicates that in the strict split scenario, training and validation data may be too distinct to enable a proper generalization.
	The AA~CDR3 setting was used for this plot.
	}
	\label{fig:performance_explain}
\end{figure}

Another surprising result is the comparably good performance of the TITAN \textit{AA CDR3} setting on unseen epitopes. 
The analysis of the attention layer in Fig.~\ref{fig:attention}a  shows that in this setting, TITAN uses an attention mask to reduce the task to a TCR classification problem. 
This should prevent the model to generalize to new epitopes. A close look at Fig.~\ref{fig:performance_explain} reveals that the best performance of the \textit{AA CDR3} model on the strict split is achieved during the first 10 to 20 epochs of training. During this time, the attention is still uniformly distributed over all input tokens, because many training epochs are required for the model to build the static attention mask described in section~\ref{ssec:attention_analysis}. 

Fig.~\ref{fig:performance_explain} also shows that  while the ROC-AUC increases continuously over the training epochs in the TCR split cross-validation, it stagnates during training on the strict split. We hypothesize that the underlying factor causing this phenomenon is the violation of the i.i.d. assumption across training and validation data.
This behavior was common during training on the strict split and again highlights the challenge for models to generalize to unseen epitopes from the sparsity of the current datasets.

While TITAN's generalization capabilities to unseen epitopes are still limited, the performance is moderately good.
This suggests that, despite their sparsity, the currently available datasets do contain enough information to enable generalization to unseen epitopes to a certain degree.
\subsection{Comparison to ImRex Model on Independent Test Set}
\label{ssec:imrex}
Moris et al. recently published ImRex~\citep{moris2020current}, an image-based generic TCR specificty prediction model, which explicitly encodes epitopes and can make predictions for unseen epitopes. As a direct comparison to ImRex, we trained TITAN with different settings on the ImRex training data and tested performance on an independent test set generated from the McPAS database (see section \ref{sec:data}). To judge both generalization capabilities to unseen TCRs and unseen epitopes, we used two different subsets of the McPAS data, one where all samples containing TCRs from the ImRex training dataset were excluded (\textit{seen epitopes} test set) and one where all samples including epitopes contained in the ImRex training dataset were removed (\textit{unseen epitopes} test set). ROC-AUC scores are compared in Tab.~\ref{tab:moriscomp}. \added{We emphasize} that this is not a cross-validation setting, i.e. we train on the full training set and record performance on an independent test set. 

We find that all of our model settings outperform ImRex on both independent test sets by a large margin. Moreover, our K-NN baseline model clearly outperforms ImRex on seen epitopes.

\begin{table}[h]
    \centering
    \caption{\textbf{Comparison of TITAN with prior work. 
    All but the ImRex model (shaded in gray) are contributions of this work. 
    Models were trained on identical data and tested on an independent test set to ensure a fair comparison.
    } 
    }
    	\begin{tabular}{ccc}
    	    & \multicolumn{2}{c}{\textbf{ROC-AUC}}\\
    	    \textbf{Model}  & \textbf{seen epitopes} & \textbf{unseen epitopes}\\
    	    \toprule
    	    \rowcolor{Gray}
    	    ImRex &
    	    0.61 &
    	    0.50 \\
    	    \hline
    	    K-NN (Baseline) &
    	    0.79 &
    	    0.37\\
    	    \hline
    	    AA CDR3 & 
    	    0.83 &	
    	    0.69\\
    	    \hline
    	    AA full &
    	    0.81 &
    	    0.64 \\
    	    \hline
    	    SMI CDR3 &
    	    0.85 &
    	    0.72\\
    	    \hline
    	    SMI full &
    	    0.86 &
    	    0.64\\
    	    \hline
    	    Pretrained & 
    	    0.79 &
    	    \textbf{0.78}\\
    	    \hline
    	    Pretrained + Aug. &
    	    0.83 &
    	    0.65 \\
    	    \hline
    	    Pretrained semifrozen &
    	    0.77 &
    	    0.69 \\
    	    \hline
    	    Pretrained semifrozen + Aug. & 
    	    \textbf{0.87} & 
    	    0.60 \\
    	    \hline
    	\end{tabular}

    	\label{tab:moriscomp}
\end{table}

Moreover, we see that the base models (\textit{AA CDR3, AA full, SMI CDR3, SMI full}) all perform better on the independent \textit{seen epitopes} test data than during 10-fold crossvalidation, while for the pretrained model settings, performance is comparable to the crossvalidation. For the \textit{unseen epitopes} test set we observe a similar trend as \added{above} in the strict split crossvalidation. All models show performances \deleted{which are} clearly better than chance, with the pretrained semifrozen model with augmentation even achieving a ROC-AUC of 0.78. However, \replaced{one needs}{we need} to keep in mind, that the sample size for the \textit{unseen epitopes} test is at only 1500, making especially high (or low) scores likely to be statistical outliers. 

\section{Discussion}

In this work, we present TITAN, a generic, bimodal, sequence-based neural network for \replaced{predicting}{prediction of} TCR--epitope binding probability that significantly \replaced{outperforms the state-of-the-art}{improves upon previous work}. 
We compare several \replaced{settings}{flavors} of TITAN that differ in their inputs \deleted{settings }for TCRs and epitopes. 
While we restrict ourselves to the TCR$\beta$ chain, we find that inputting the complete variable TCR sequence boosts performance compared to scenarios where only the CDR3 region is used. 

Notably, we are, to the best of our knowledge, \added{the} first to formulate TCR-epitope binding prediction as \added{a} subclass of the commonly investigated task of \added{predicting} compound-protein-interaction\deleted{prediction}.
The concomitant change of representing epitopes as SMILES (an atomic, more granular representation) instead of amino acid sequences improves the predictive power of TITAN.
This reformulation not only enables data augmentation --- by exploiting the non-uniqueness of SMILES~\citep{Bjerrum2017} --- to enrich the training data, but also positions us to leverage large-scale datasets \added{ such as} BindingDB for pretraining. 
Pretrained TITAN models achieve considerably higher scores than all TITAN base models. 
Freezing the weights of the epitope input channel coupled with SMILES augmentation gives us the best performance with a mean ROC-AUC of $0.868 \pm 0.005$ on the TCR split. 

To assess model performance in a more rigorous fashion, we designed a K-NN classifier based on the Levenshtein distance as a baseline model. 
Surprisingly, this simple model achieves a mean ROC-AUC score of $0.779 \pm 0.007$ on the TCR split crossvalidation.
This model surpassed \added{the} performance of complex neural networks from previous publications~\citep{moris2020current} and is only outperformed by our pretrained TITAN \replaced{models}{settings}, which highlights the importance of including relevant baseline models. 

As a final assessment of model performance, we compare TITAN to the current state of the art for general TCR specificity prediction, ImRex. To ensure fair comparison, we train our models on the ImRex training data and evaluate \added{the} performance on an independent test set derived from a different database. We demonstrate that both pretrained and base TITAN models, as well as the K-NN baseline, \replaced{outperform}{beat} ImRex by a large margin. The best result is again achieved by \added{the} pretrained semifrozen augmented TITAN model, with a ROC-AUC of 0.87 on epitopes included in the training data. 

Finally, the main challenge for generic TCR--epitope interaction prediction remains the generalization to unseen epitopes. Here, we report two major breakthroughs. First, we demonstrate that TITAN exhibits moderate performance on unseen epitopes, where the best TITAN \replaced{model}{setting} achieves a ROC-AUC of $0.62 \pm 0.05$ on a strict 10-fold crossvalidation split, and ROC-AUC of 0.78 on an independent test set of unseen epitopes. Second, by dissecting the attention heatmaps we were able to identify a possible explanation for \added{the observed} poor unseen epitope generalization \added{capabilities}. We demonstrate that TITAN reduces the general TCR--epitope prediction task to the simpler task of TCR classification, by internally treating the epitopes as classes instead of focusing on their properties. This shortcut could present a general problem for sufficiently complex models, as long as the extreme sparsity of sampling of the epitope sequence space is not remedied.
Until then, \added{our future endeavors might include using an enriched training dataset consisting of TCR-epitope pairs as well as compound-protein interaction pairs from BindingDB, or further improving the amino acid and SMILES embeddings by training on diverse databases (UniProt~\cite{uniprot2020}, chembl~\cite{chembl2016}).}
\added{In general,} our results indicate that approaches with a focus on leveraging transfer learning techniques 
to enrich the input data space may be \replaced{promising}{ a promising approach toward} \replaced{to tackle}{improved generalization to} the daunting task of unseen epitope-TCR specificity prediction. 

\newpage
\section{Acknowledgements}
The authors acknowledge funding from the European Union's Horizon 2020 research and innovation programme under the Marie Sklodowska-Curie grant agreement No. 813545 and 826121.
We thank Matteo Manica and Joris Cadow for building up much of the model architecture used, as well as for many helpful discussions.

\pagestyle{plain}
\pagenumbering{arabic}

\bibliographystyle{abbrvnat}
\begin{small}
\bibliography{references} 
\end{small}

\end{document}